\documentclass{article}

\usepackage{arxiv}

\usepackage[utf8]{inputenc} 
\usepackage[T1]{fontenc}    
\usepackage{hyperref}       
\usepackage{url}            
\usepackage{booktabs}       
\usepackage{amsfonts}       
\usepackage{nicefrac}       
\usepackage{microtype}      
\usepackage{lipsum}
\usepackage{subcaption}
\usepackage{multirow}
\usepackage{makecell}
\usepackage{hyperref}
\usepackage{graphicx}

\title{Compact Ring Resonator Enhanced Silicon-MSM Photodetector in SiN-on-SOI Platform}

\author{
  Avijit Chatterjee\thanks{Corresponding author: avijit@iisc.ac.in}  \\
  Center for Nano Science and Engineering\\Indian Institute of Science Bangalore\\
  \texttt{avijit@iisc.ac.in} \\
   \And
 Saumitra Sam \\
  Molecular Biophysics Unit\\ Indian Institute of Science Bangalore\\
   \And
 Sujit Kumar Sikdar\\
  Molecular Biophysics Unit\\ Indian Institute of Science Bangalore\\
   \And
 Shankar Kumar Selvaraja \\
  Center for Nano Science and Engineering\\ Indian Institute of Science Bangalore\\
  \texttt{shankarks@iisc.ac.in} \\
}

\begin{document}
\maketitle

\begin{abstract}
We present a compact on-chip resonator enhanced silicon-MSM photodetector in $850$ $nm$ wavelength band for communication and lab-on-chip bio-sensing applications. We report the highest responsivity of 0.81 A/W for a 5 $\mu m$ long device. High responsivity is achieved by integrating the detector in a silicon nitride ring resonator. The resonance offers 100X responsivity improvement over a single-pass photodetector due to cavity enhancement. We also present a detailed study of the high-speed response of the cavity and single-pass detector. We report an electro-optic bandwidth of 7.5 GHz measured using a femtosecond optical excitation. To the best of our knowledge, we report for the first time silicon nitride resonator integrated Si-MSM detector in SiN-SOI platform.
\end{abstract}


\section{Introduction}
Exponential growth of internet traffic pose the requirement for ultra-high data rate communication links. Integrated optical interconnect is a promising candidate to offer high-speed, scalable, and affordable solution in such situations\cite{Miller2009,Taubenblatt2012,Tatum2015}. In an optical interconnect, one of the key components is a high-speed photodetectors\cite{soref2006past,IEEE400g,Taubenblatt2012,Miller2009,Tatum2015,Kachris2012a}.
Silicon photonics based on-chip interconnects with highly efficient Ge/III-V photodetectors is already a very matured, and commercially used platform\cite{Chen:09, vivien2012zero,sheng2010ingaas,liu2010iii,spuesens2012compact}. For short reach optical interconnects at 850 nm wavelength, silicon photodetector is also explored in addition to Ge and III-V photodetectors\cite{chen2018integration,ciftcioglu2010850,hsu2003high}. The reason is that the silicon photodetectors have higher responsivity than Ge/III-V at 850 nm wavelength and CMOS compatible fabrication. However, silicon has lower bandwidth than Ge or III-V, which limits its application for high-speed interconnects. To improve the bandwidth of silicon photodetectors, recently thin silicon-on-insulator (SOI) platform is explored \cite{gao2017photon,PourFard2017,Li2015,Lee2016}. For a conventional top illuminated photodetector, thin silicon reduces the cross-section area and improves the RC time limited bandwidth resulting in sub-picosecond response time \cite{liu1994140}. However, reduced cross-section area decreases responsivity \cite{liu1994140}. To obtain higher responsivity without degrading bandwidth, waveguide photodetector configuration is a better option than a top illuminated photodetector. An on-chip lateral silicon pin photodetector integrated with SiN waveguide on SiN-SOI platform with higher responsivity ($0.44$ $A/W$) and large bandwidth ($14$ $GHz$) was demonstraed in \cite{chatterjee2019high}.  Waveguide integrated photodetector also offers opportunity for on-chip passive optical device integration, such as wavelength selective devices that are essential for next-generation optical datacom and lab-on-chip bio-sensing application. The responsivity and bandwidth of an integrated silicon photodetector need to improve further, to attain the 400 Gbps road map for short-reach optical datacom \cite{IEEE400g}. Therefore, the challenge is to increase the responsivity of the photodetector without increasing the physical size to maintain high bandwidth. To achieve high responsivity and bandwidth, one of the solutions is a cavity enhanced photodetector \cite{neudeck1998selective,schaub1999resonant,cheng2005silicon,alloatti2016resonance,casalino2018design,Huang:18,nozaki2016photonic,zhou2016enhanced}. In the cavity enhanced photodetector, the photodetector is placed inside a resonating cavity. Since the cavity offers optical field enhancement, one could achieve higher responsivity \cite{schaub1999resonant,neudeck1998selective,cheng2005silicon,Huang:18}. In addition, the cavity size can be made compact to achieve large bandwidth. In literature various cavity configuration such as Bragg mirror \cite{schaub1999resonant}, photonic crystal \cite{nozaki2016photonic,nozaki2016photonic}, ring resonator\cite{song2014microring}, and metal reflector \cite{collin2003resonant} were used to realize the cavity enhanced photodetector. 
\\
In this paper, we demonstrate silicon nitride (SiN) ring resonator enhanced silicon metal-semiconductor-setal (MSM) photodetector in the 850 nm wavelength band on a SiN-SOI platform. In recent years, SiN has emerged as a versatile platform for integrated photonic circuit for broad spectrum of applications \cite{subramanian2013low,Chen:18,daldosso2004comparison}. The emergence is due to broad wavelength transparency and CMOS compatibility. Since SiN is a dielectric material, active functionality such as light modulation, generation and detection requires heterogeneous or monolithic integration of suitable material. In this work, we demonstrate CMOS compatible monolithic integration of Si photodetector. In particular, we demonstrate waveguide Si-MSM photodetector integrated in a SiN ring resonator. Unlike doped junction photodector, MSM photodetector offers high speed and lower dark current \cite{liu1994140}. In addition, MSM fabrication is not as demanding as doped junction. In a doped junction photodetector, the size of the photodetector depends on the control of junction alignment and dopant diffusion control \cite{jeyaselvan2018lateral}. However, in a MSM detector fabrication, the metal electrodes can be lithographically defined with sub-50 nm alignment accuracy. In this paper, we focus on realizing high-speed and high responsivity waveguide integrated Si photodetector integrated with SiN ring resonator on a single platform for 850 nm based short reach optical interconnects.

\begin{figure}
\centering\includegraphics[scale=0.4]{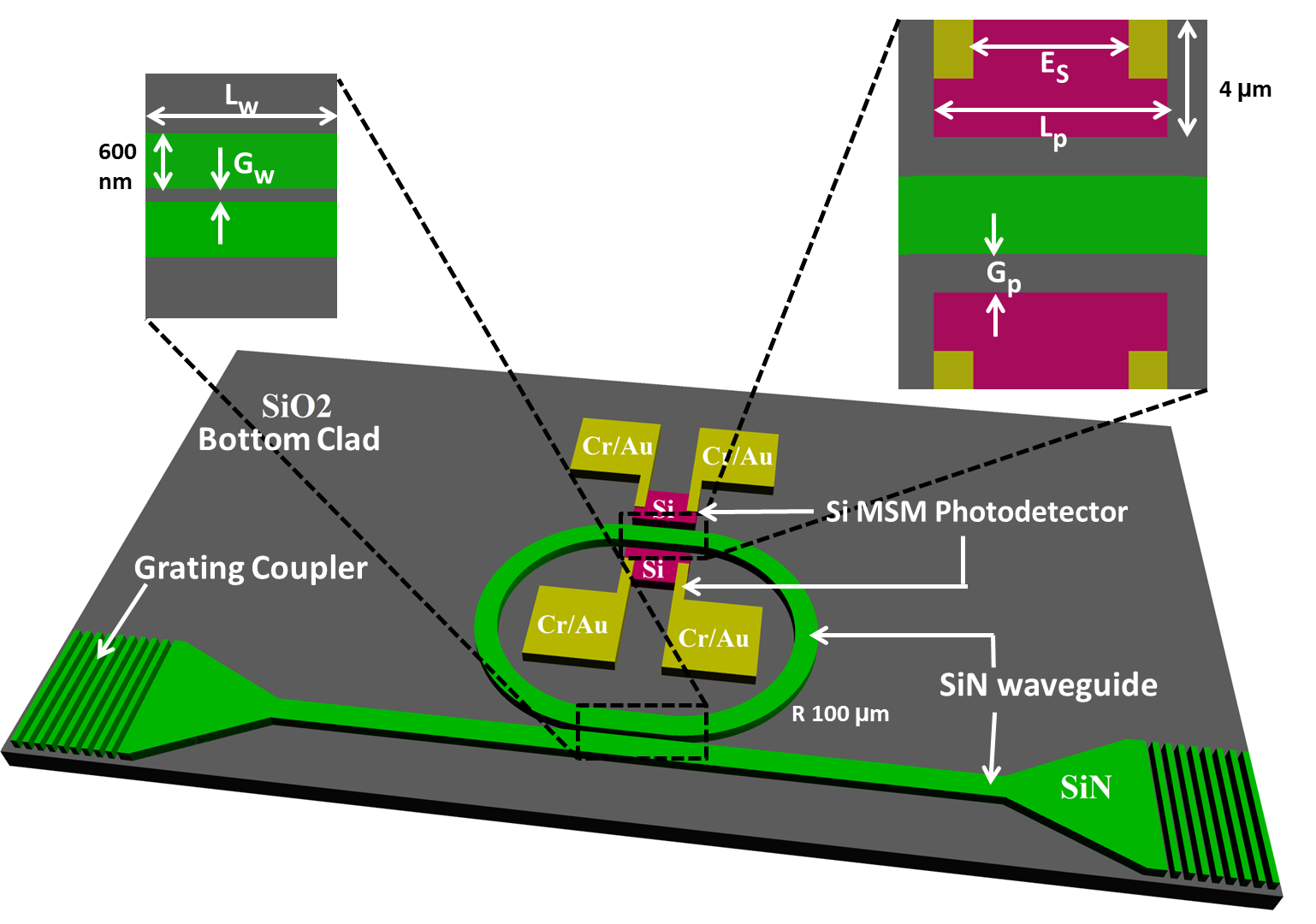}
\caption{Schematic of a ring coupled silicon-MSM photodetector (RCPD)}
\label{fig:model}
\end{figure}

\begin{figure}
\centering\includegraphics[scale=0.6]{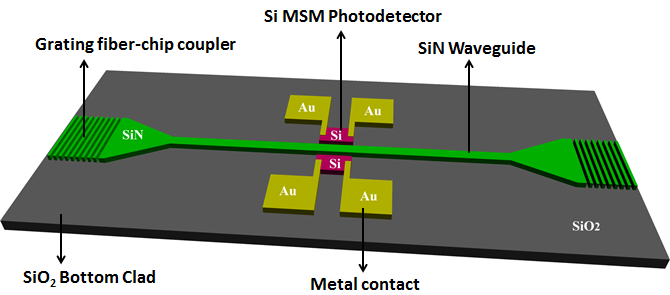}
\caption{Schematic of a waveguide coupled silicon-MSM photodetector (WGPD)}
\label{fig:wgpd}
\end{figure}

\section{Design and simulation}
 
Figure \ref{fig:model} shows a schematic of the proposed ring coupled silicon MSM photodetector (RCPD). The proposed device comprises of a race-track SiN ring resonator and Si-MSM photodetector. A Si-MSM photodetector is integrated with the SiN ring resonator in such a way that the circulating optical field inside the SiN ring resonator evanescently couples into the Si photodetector. The gap ($G_{p}$) and the interaction length ($L_{p}$) between the photodetector and SiN waveguide determine power coupling efficiency and responsivity subsequently. The SiN waveguide circuit was designed to operate in single-mode (TE) regime at 850 nm. Accordingly, based on modal calculations, SiN waveguide width and thickness of 600 nm and 220 nm were chosen, respectively. Light coupling between an optical fiber and the waveguide is achieved using a grating coupler and adiabatic taper \cite{chatterjee2019high}. The grating were 220 nm etched with a grating period of 650 nm. 
  
In addition to RCPD, a conventional waveguide integrated Si-MSM photodetector (WGPD) as shown in Fig. \ref{fig:wgpd} is also realized to compare the performance with the RCPD. The configuration and design dimensions of RCPD and WGPD are made identical in order to evaluate the responsivity enhancement of RCPD over WGPD.

\begin{figure}[h]
\begin{subfigure}{0.45\textwidth}
\centering
\includegraphics[scale=0.3]{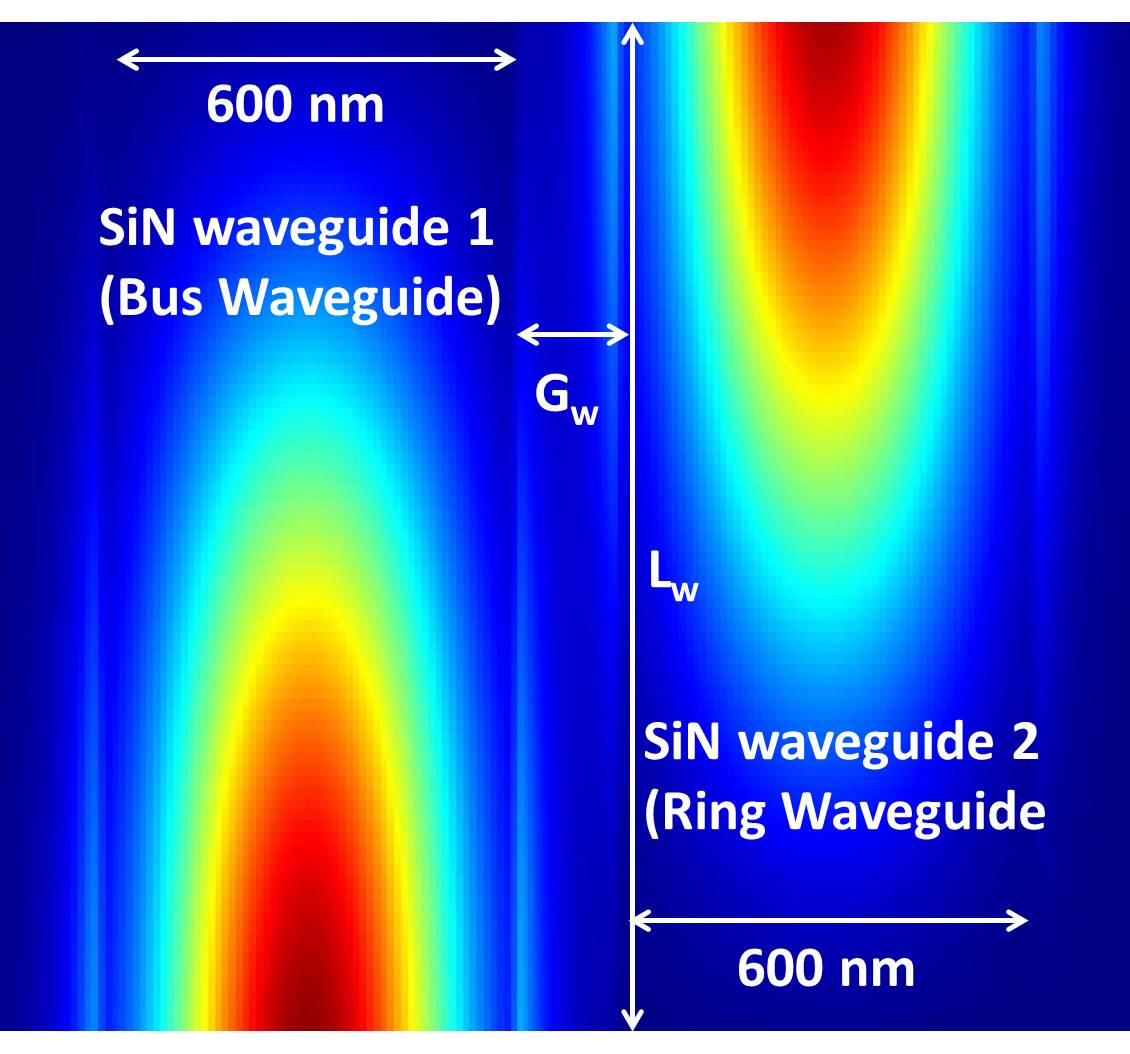} 
\vspace{-0.2cm}
\end{subfigure}
\begin{subfigure}{0.5\textwidth}
\centering
\includegraphics[scale=0.07]{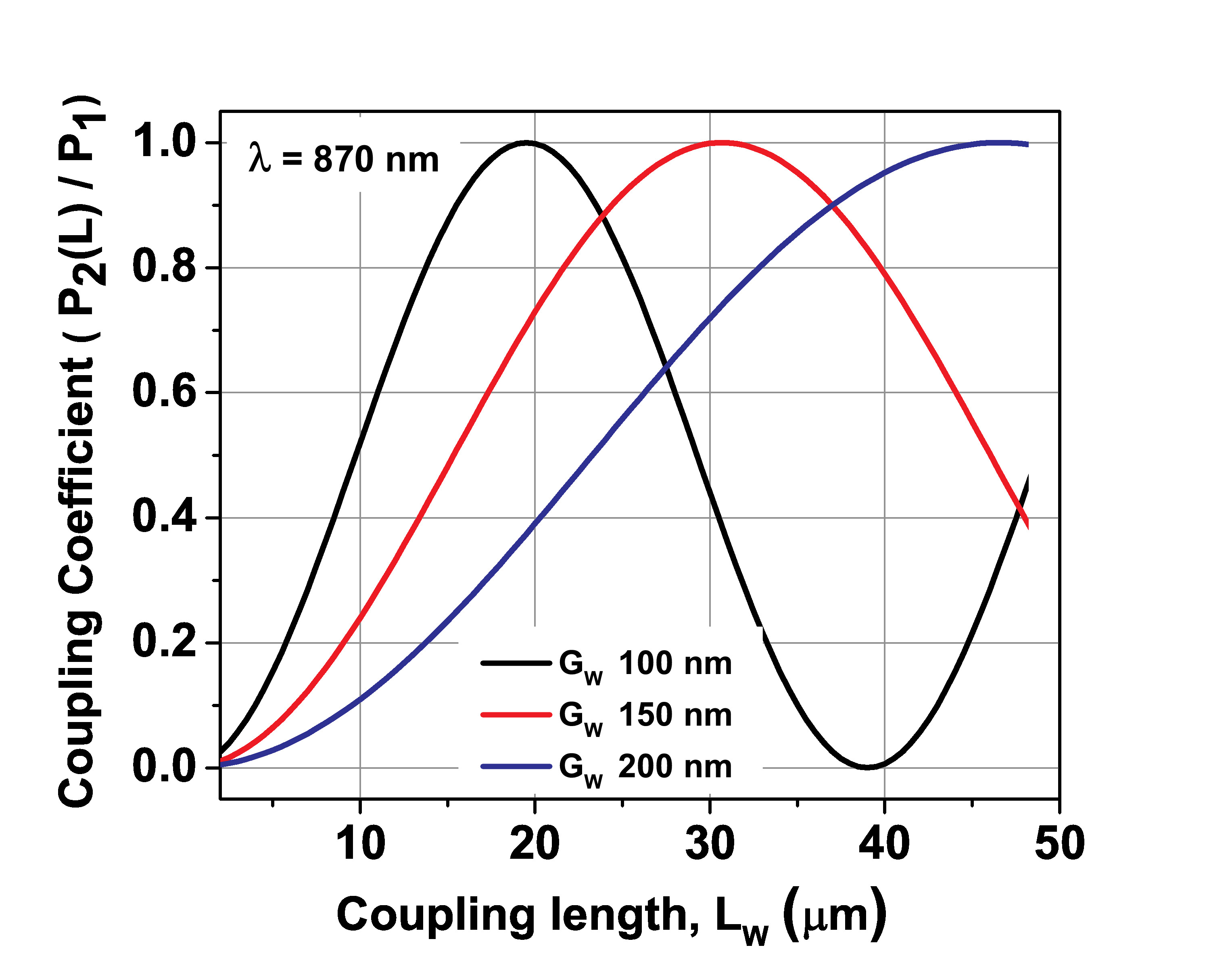}
\vspace{-0.1cm}
\end{subfigure}
\vspace{-0.3cm}
\caption{Power coupling between a bus waveguide and a ring waveguide, a) Electric field distribution along the length of the directional coupler and b) coupling efficiency as a function of coupling length ($L_{W}$) for varying gap ($G_{W}$) }
\label{fig:directional_coupler}
\end{figure}

A racetrack ring resonator is used as a cavity. The design parameters are optimized using FDTD simulations \cite{lumerical2016solutions}. The gap between the ring resonator and the bus waveguide ($G_{w}$), and the coupling length ($L_{w}$) are optimized to achieve required power coupling into the ring. Figure \ref{fig:model} illustrates a racetrack ring resonator with integrated Si-MSM detector. The waveguide width of the ring and bus waveguide are kept identical at 600 nm. A ring radius of 100 $\mu m$ is chosen to reduce the bend loss. The resonator is operated in a critically coupled regime to achieve sufficient power coupling into the cavity. To achieve critical coupling, the cavity loss is estimated from the propagation loss measurement. A propagation loss of  2.5 dB/mm is obtained from cut-back method. The propagation loss is dominated by the sidewall roughness form the fabrication process. By optimizing the fabrication process, one could achieve one-order lower loss \cite{daldosso2004comparison}. Based on the estimated cavity round trip loss, required coupling efficiency is obtained by optimizing $G_{w}$ and $L_{w}$  \cite{solutionsversion}. Figure \ref{fig:directional_coupler} shows the power coupling in a directional coupler and the effect of $L_{W}$ and $G_{W}$ between the waveguides on  the coupling efficiency. Based on the simulation and fabrication limitations, we used $G_{w}=150$ nm and $L_{w}=6$ $\mu m$ as the optimum design parameters.  
   
In addition to power coupling into the ring, coupling from SiN ring waveguide into Si photodetector needs to be optimized to achieve optimum responsivity enhancement. It is essential to maintain a circulating optical field in the ring resonator. The power enhancement in the cavity depends on the absorption in the Si-MSM detector and loss in the SiN ring waveguide. An excess optical power loss in the cavity will diminish the circulating power in the resonator resulting in resonance degradation. Since the field enhancement in the cavity is crucial in achieving responsivity enhancement, the power coupling into the cavity and the detector needs careful optimization. Considering the cavity loss, we have chosen $G_{p}$ as $100$ nm and varied $L_{p}$ in the range of 3-6 $\mu m$ to study the effect of coupling and cavity loss.

\begin{figure}[!h]
\centering\includegraphics[scale=0.2]{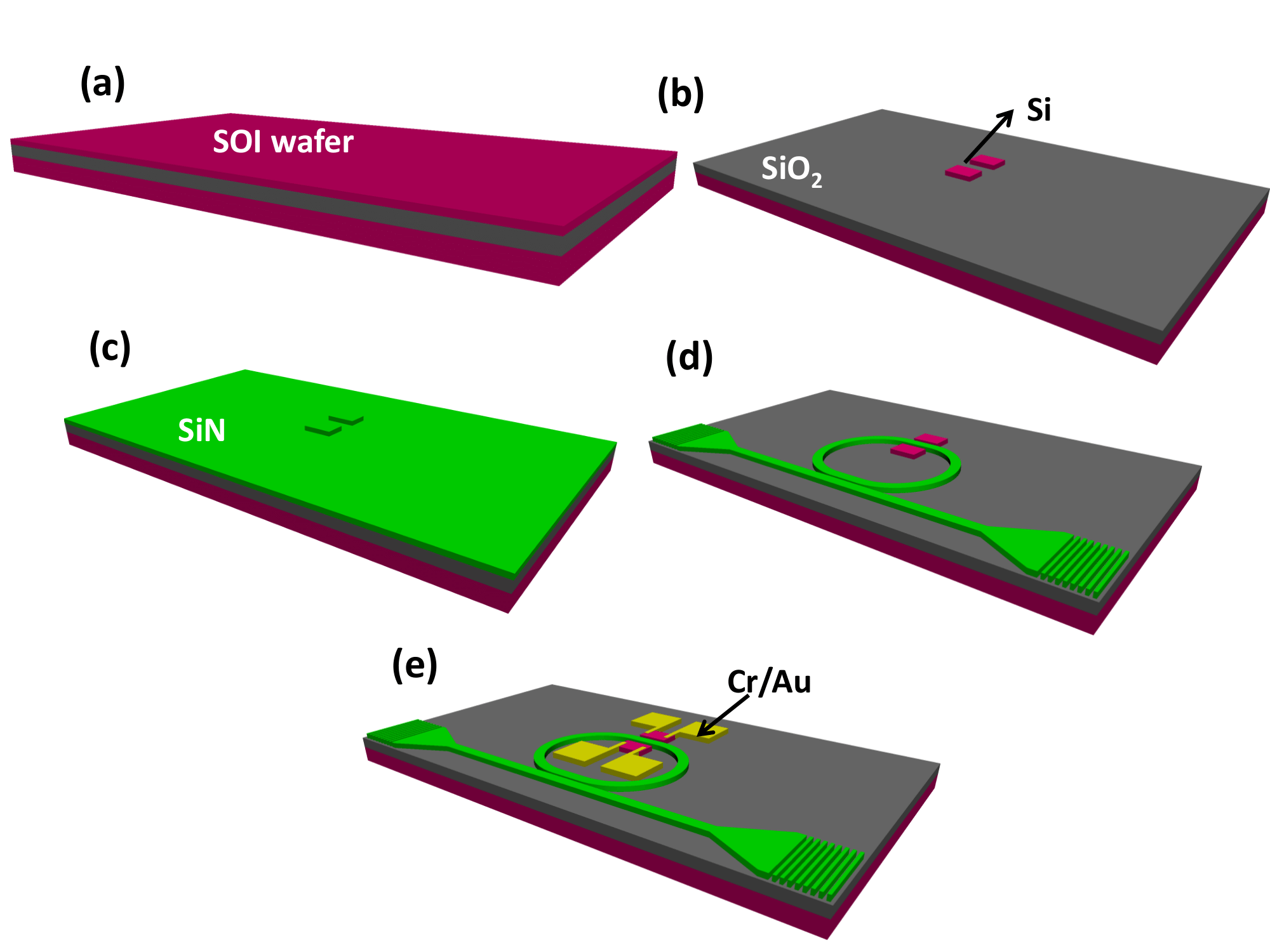}
\caption{Fabrication process flow overview; (a) Cleaned SOI wafer, (b) silicon patterned by e-beam lithography and dry etching, (c) 50 nm PECVD $SiO_{2}$ and 220 nm LPCVD SiN deposited, (d) SiN is patterned by e-beam lithography and dry etching, (e) Chrome/Gold contacts made by e-beam lithography, e-beam evaporation and lift off}
\label{fig:fabflow}
\end{figure}

\section{Fabrication}

\begin{figure}[!h]
\centering\includegraphics[scale=0.5]{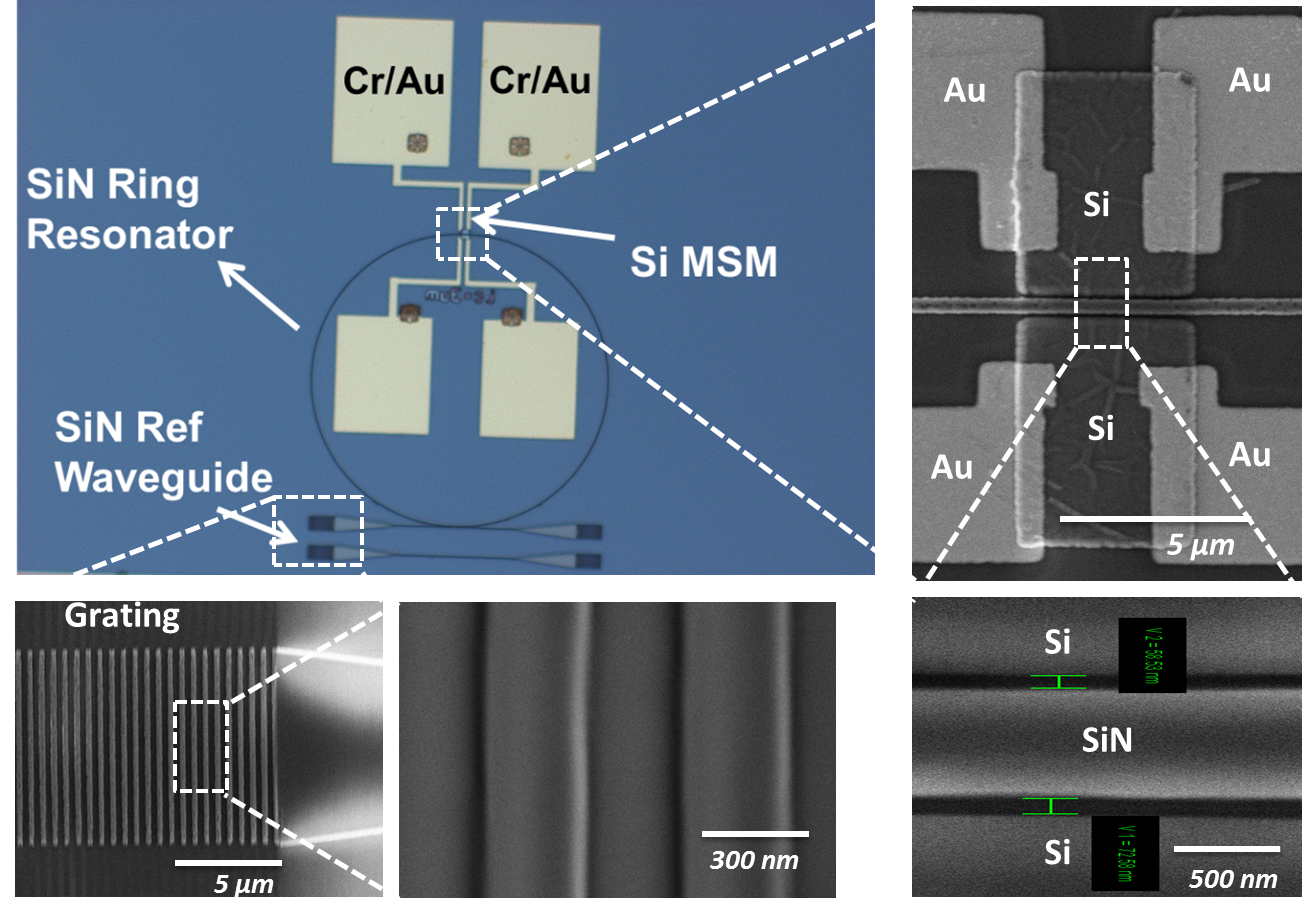}
\caption{Optical and scanning electron microscope image of a fabricated ring coupled Si-MSM detector.}
\label{fig:fabsem}
\end{figure}

The device is fabricated on a SOI wafer with 220 nm thick silicon device layer on a 2 $\mu m$ thick buried oxide. An overview of the fabrication process flow is illustrated in Fig. \ref{fig:fabflow}. After standard wafer clean, Si was patterned using electron-beam lithography (EBL) followed by inductively coupled reactive-ion etching (ICP-RIE) forming islands of Si (Fig. \ref{fig:fabflow}b). A 50 nm thick liner $SiO_{2}$ followed by 220 nm thick SiN by PECVD and LPCVD, respectively. The liner oxide acts as a passivation layer, and etch-step layer for SiN. The SiN layer was then patterned using EBL and dry etching of 220 nm thick SiN by ICP-RIE to define the waveguide, the grating coupler, and the ring resonators in a single patterning step. The SiN over the detector region is removed during the etch process stopping on the liner oxide. Finally, chrome-gold electrical contacts were formed on Si using a lift-off process. Figure \ref{fig:fabsem} shows optical and scanning electron microscope image of one of the fabricated cavity Si-MSM photodetector.

\section{Results and discussion}

The fabricated devices were characterized for static and dynamic optoelectronic characteristics. The static characteristics were measured using source measurement units while the dynamic response was measured using a pulsed laser and a high-speed oscilloscope.

\subsection{I-V characteristic of ring resonator enhanced silicon-MSM photodetector}

Electrical response was measured for Si-MSM photodetector under dark and light condition. An LED source with the wavelength range of 860-900 nm was used for the measurement. Light from an LED source is coupled using an optical fiber into the waveguide through grating coupler. The spectral response of resonator was captured by an optical spectrum analyzer (OSA). To bias the Si-MSM photodetector and measure the output current, source measuring unit (Keithley 2401) and DC probes were used.
 
\begin{figure}
\centering\includegraphics[scale=0.1]{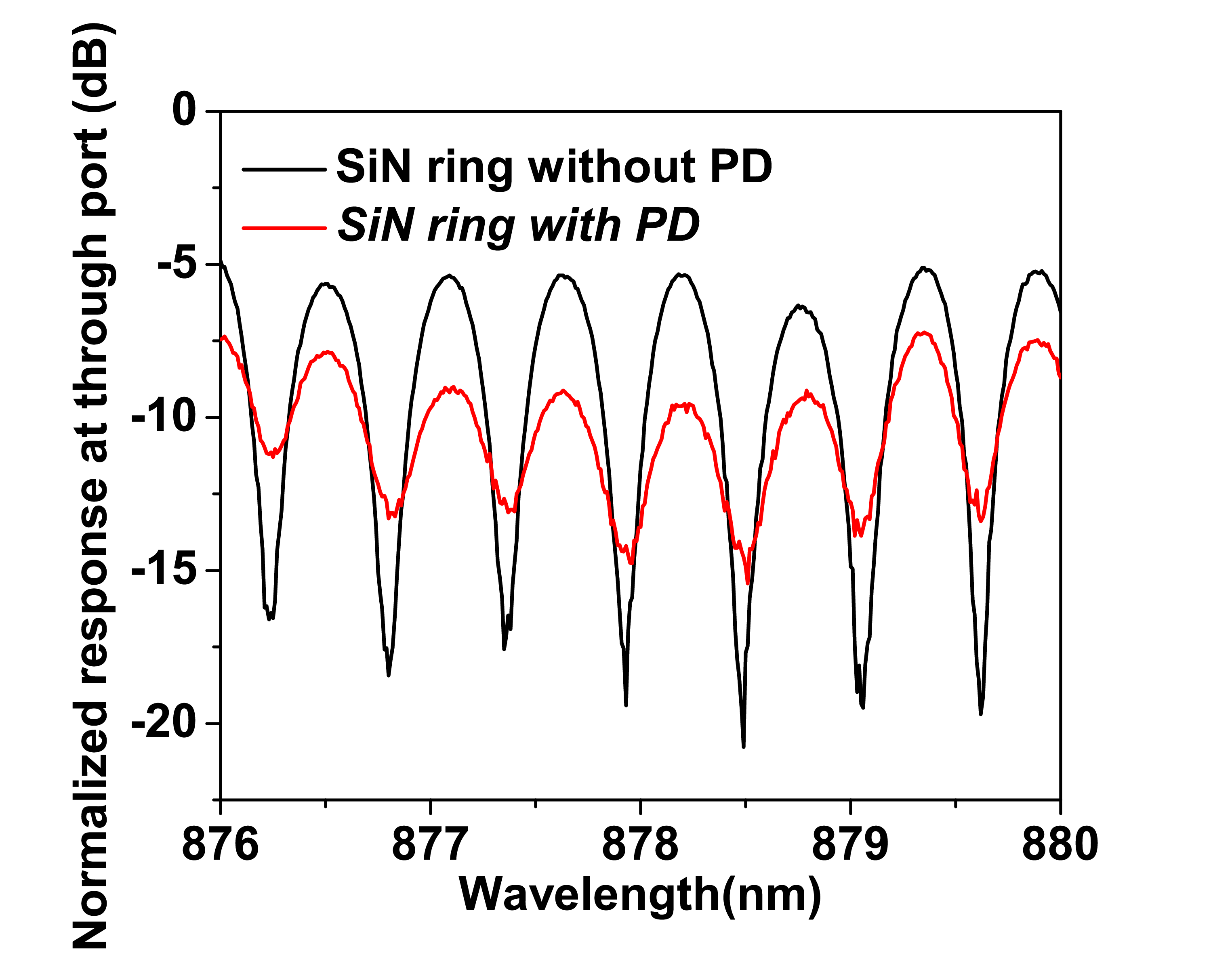}
\caption{Normalized through port ring resonator spectral response; with and without Si-MSM integration.}
\label{fig:ringresponseboth}
\end{figure}

\begin{figure}[h]
\begin{subfigure}{0.45\textwidth}
\centering
\includegraphics[width=1\linewidth]{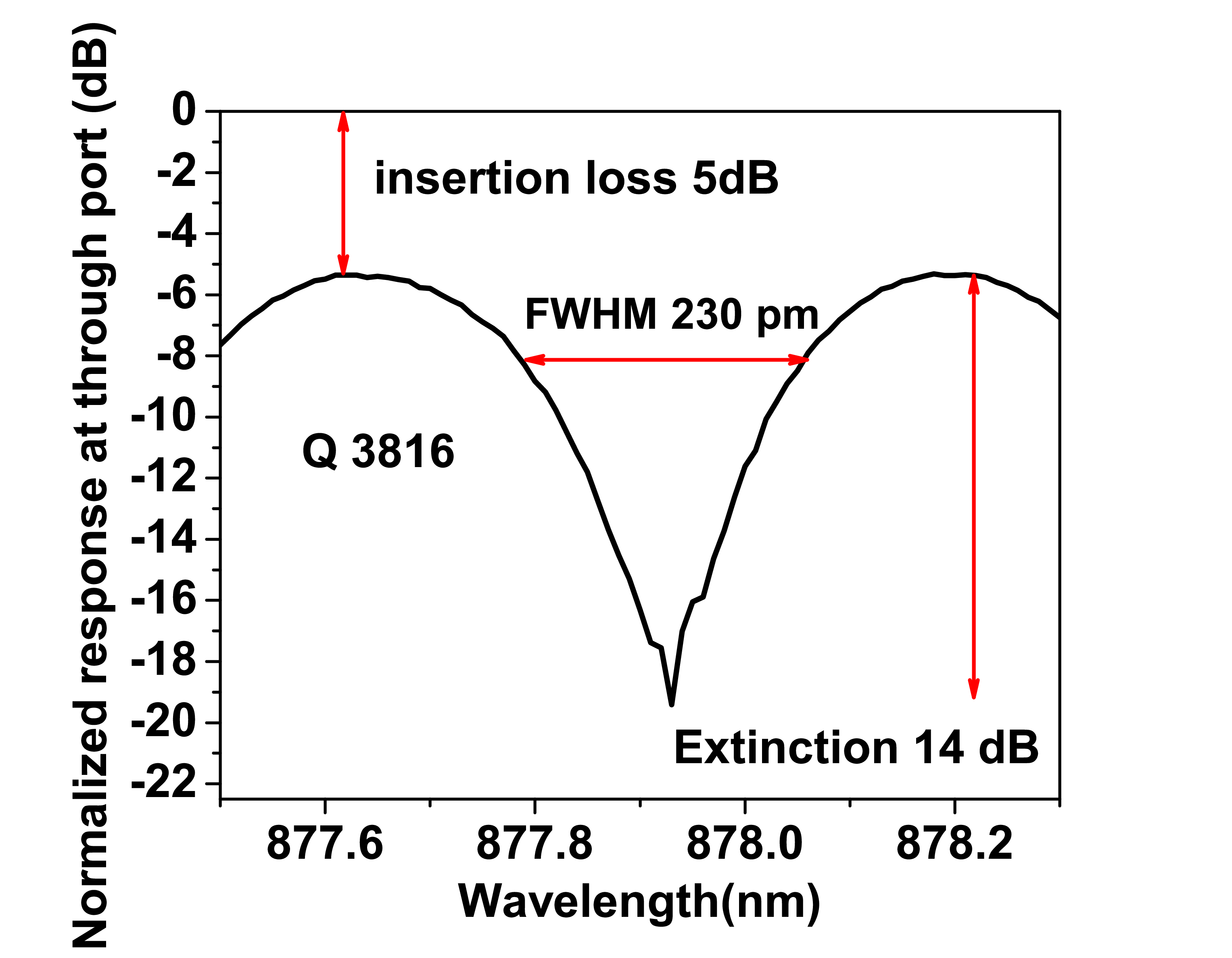} 
\vspace{-0.7cm}
\caption{Without Si-MSM}
\label{fig::ring}
\end{subfigure}
\begin{subfigure}{0.45\textwidth}
\centering
\includegraphics[width=1\linewidth]{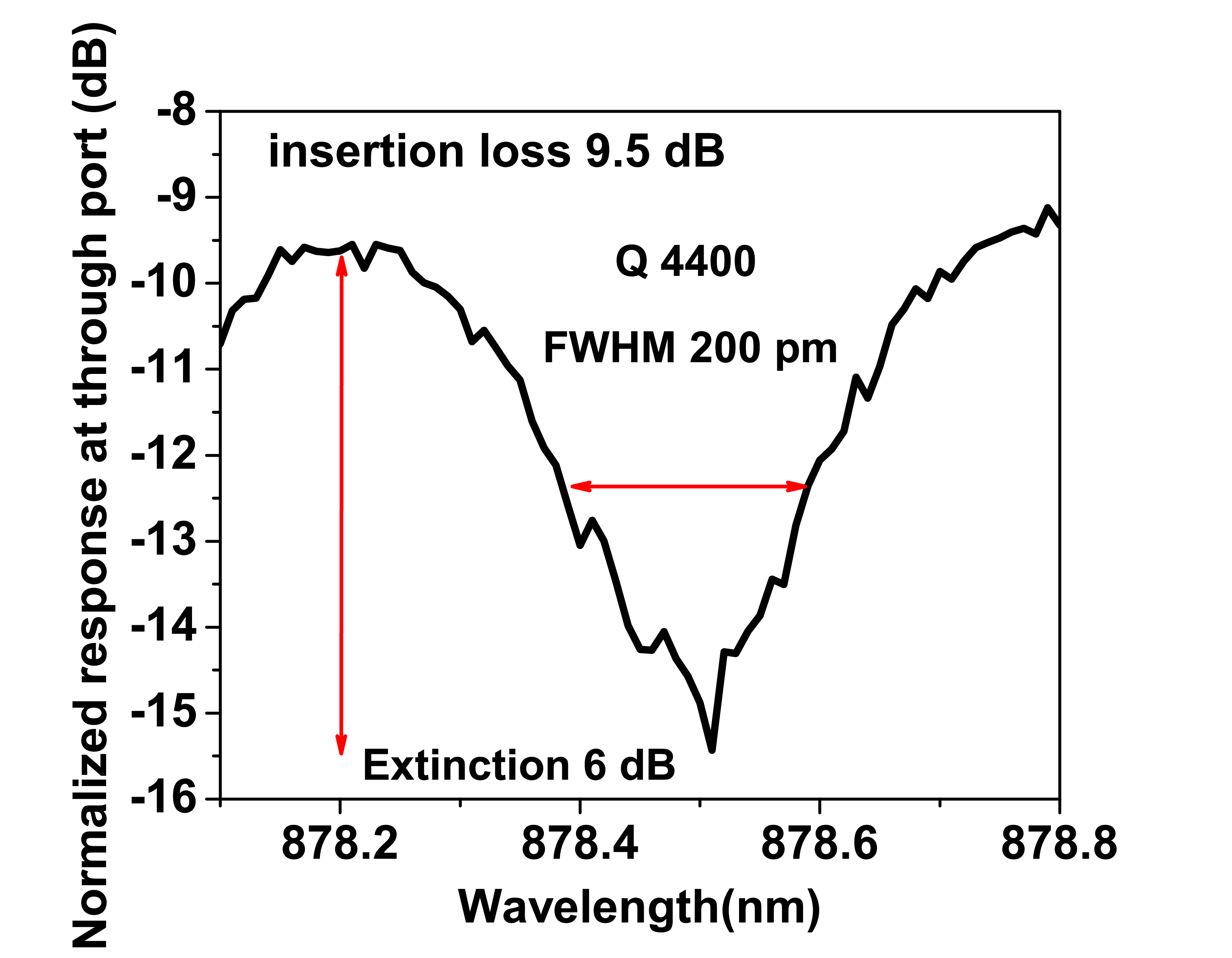}
\vspace{-0.7cm}
\caption{With Si-MSM} 
\label{fig:ringmsm}
\end{subfigure}
\vspace{-0.3cm}
\caption{Normalized isolated single through port spectral resonance.}
\label{fig:ring_nomsm_msm}
\end{figure}

 Figure \ref{fig:ringresponseboth} shows the normalized transmission response of the fabricated ring resonator with and without Si-MSM photodetector. The obtained spectrum is normalized with a reference waveguide. A coupling efficiency of 14 dB/coupler at 860 nm and a 1-dB bandwidth of 12 nm was measured from the reference waveguide. The spectral response of the ring with and without MSM photodetector clearly shows the effect of Si-MSM. We observe an increase of $\approx$4 dB in insertion and $\approx$8 dB in extinction due to Si-MSM in the ring cavity (Fig. \ref{fig:ring_nomsm_msm}). In addition, a marginal increase in the quality factor due to the detector (Fig. \ref{fig:ring_nomsm_msm}b). The increase in the quality factor can be attributed to ring response moving towards critical coupling from an over-coupled regime.

%
%

Figure \ref{fig:IV} (a) summarizes I-V characteristics of the Si-MSM photodetector with and without optical excitation. Figure \ref{fig:IV} (a) and (b) also shows a comparison between the ring coupled detector (RCPD) and waveguide coupled detector (WGPD). The RCPD shows lower dark current of  $\approx$0.8 nA compared to 10 nA measured at -5 V bias for a WGPD of same detector dimension. Though the length of the detector is the same for WGPD and RCPD, RCPD measures lower dark current. This is due to a minor difference in the contact configuration between the two detector configuration. In RCPD, the electrodes on the outer and inner side of the ring have different spacing which substantially reduces the dark current, while in a WGPD the electrodes are equally placed on both sides of the waveguide. Furthermore, the asymmetry in the IV characteristics is due to asymmetric contacts. We measure a photocurrent of 1 $\mu$A for an RCPD and 30 nA for a WGPD with a bias of -5 V. Unlike WGPD, due to cavity enhancement RCPD shows three orders of magnitude increase in light current under illumination.

\subsection{Responsivity estimation}

The responsivity of the RCPD is estimated from the optical power in the ring waveguide. First, power coupled into the bus waveguide is calculated by deducting the fiber to grating coupler loss. The power coupled from the waveguide into the ring is estimated from the directional coupler (Fig. \ref{fig:directional_coupler}). Further, the enhancement due to the cavity is also considered in responsivity calculation. 

\begin{figure}[h]
\begin{subfigure}{0.45\textwidth}
\centering
\includegraphics[width=1\linewidth]{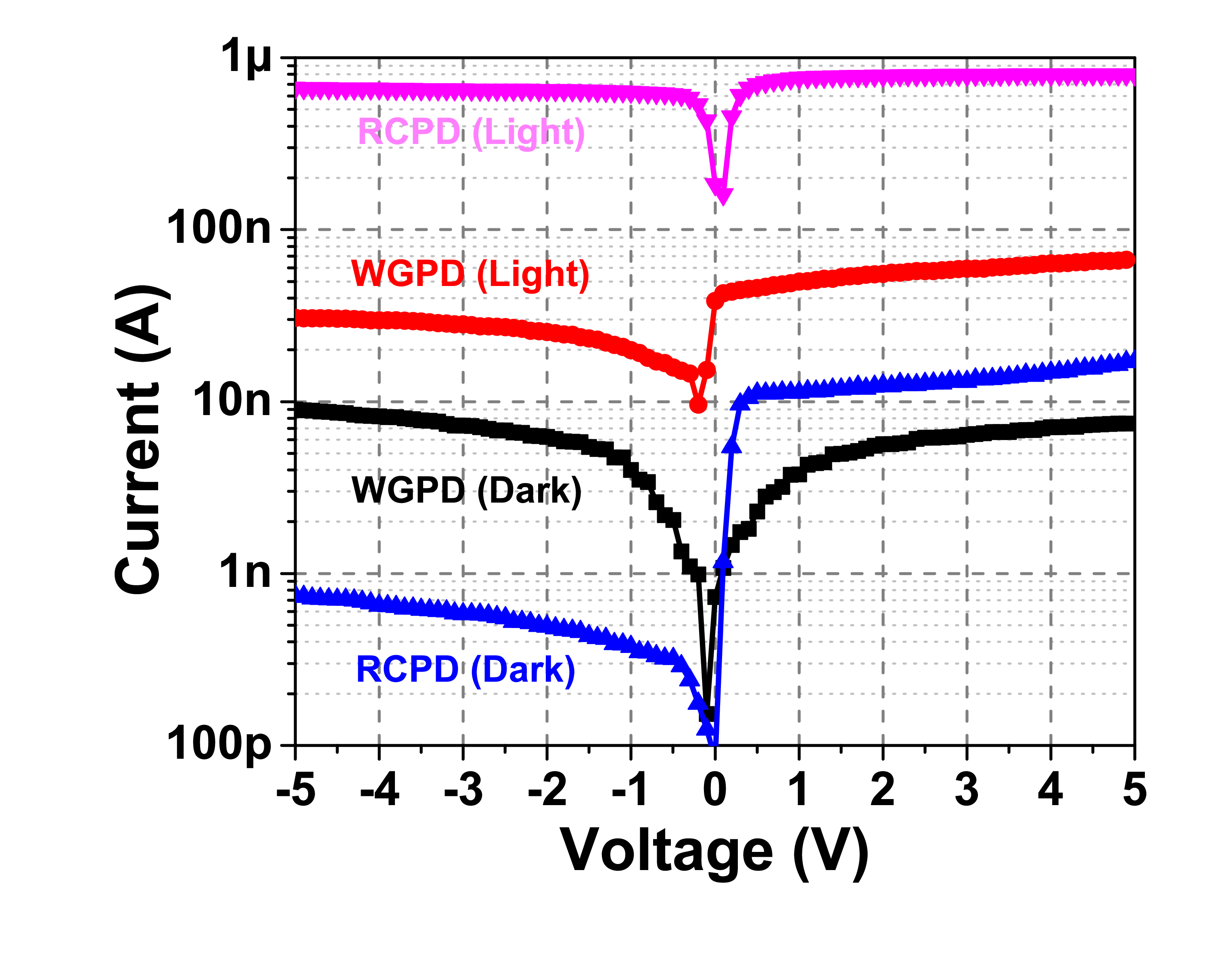} 
\vspace{-0.7cm}
\caption{Current Vs Voltage}
\label{fig::IVwgrc}
\end{subfigure}
\begin{subfigure}{0.45\textwidth}
\centering
\includegraphics[width=1\linewidth]{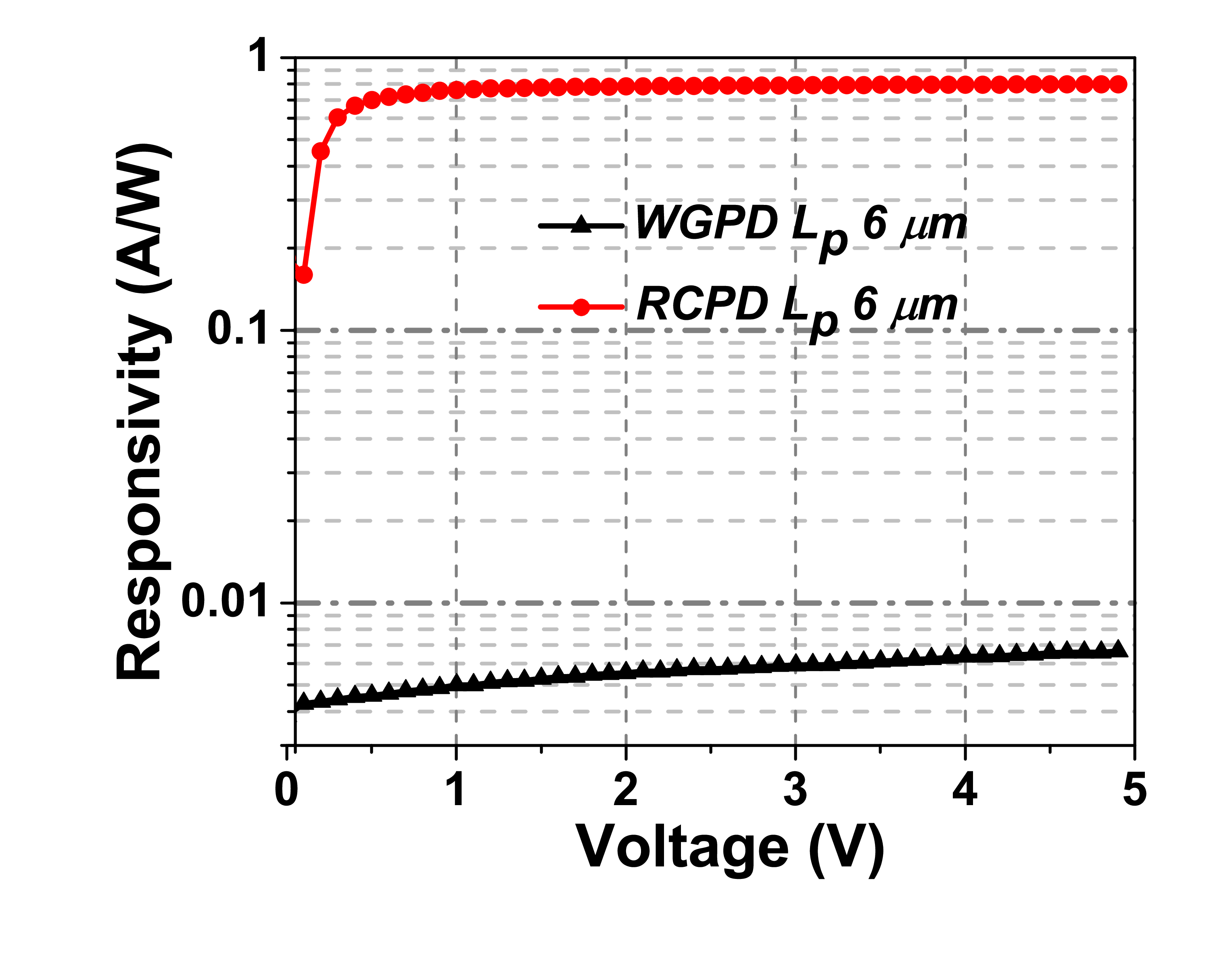}
\vspace{-0.7cm}
\caption{Responsivity Vs Voltage}
\label{fig:Rwgrc}
\end{subfigure}
\vspace{-0.3cm}
\caption{IV characteristics and responsivity of waveguide coupled (WGPD) and ring coupled (RCPD) Si-MSM photodetector with $L_{p}$ = 6 $\mu m$, $G_{w}$ = 150 nm, $L_{w}$ = 6 $\mu m$ }
\label{fig:IV}
\end{figure}

The responsivity of the RCPD and WGPD is calculated from the power coupling, and IV characterizes. We measure a responsivity of 0.79 A/W at 5 V for an RCPD, which is two orders higher than the WGPD (Fig. \ref{fig:IV}b). Improvement in responsivity is attributed to the enhanced optical field inside the ring and absorption of optical power \cite{Huang:18,casalino2018design}. We observe a marginal increase in the responsivity due to higher field between the electrodes that efficiently sweeps the photo-generated carriers. 

\begin{figure}
\centering\includegraphics[scale=0.1]{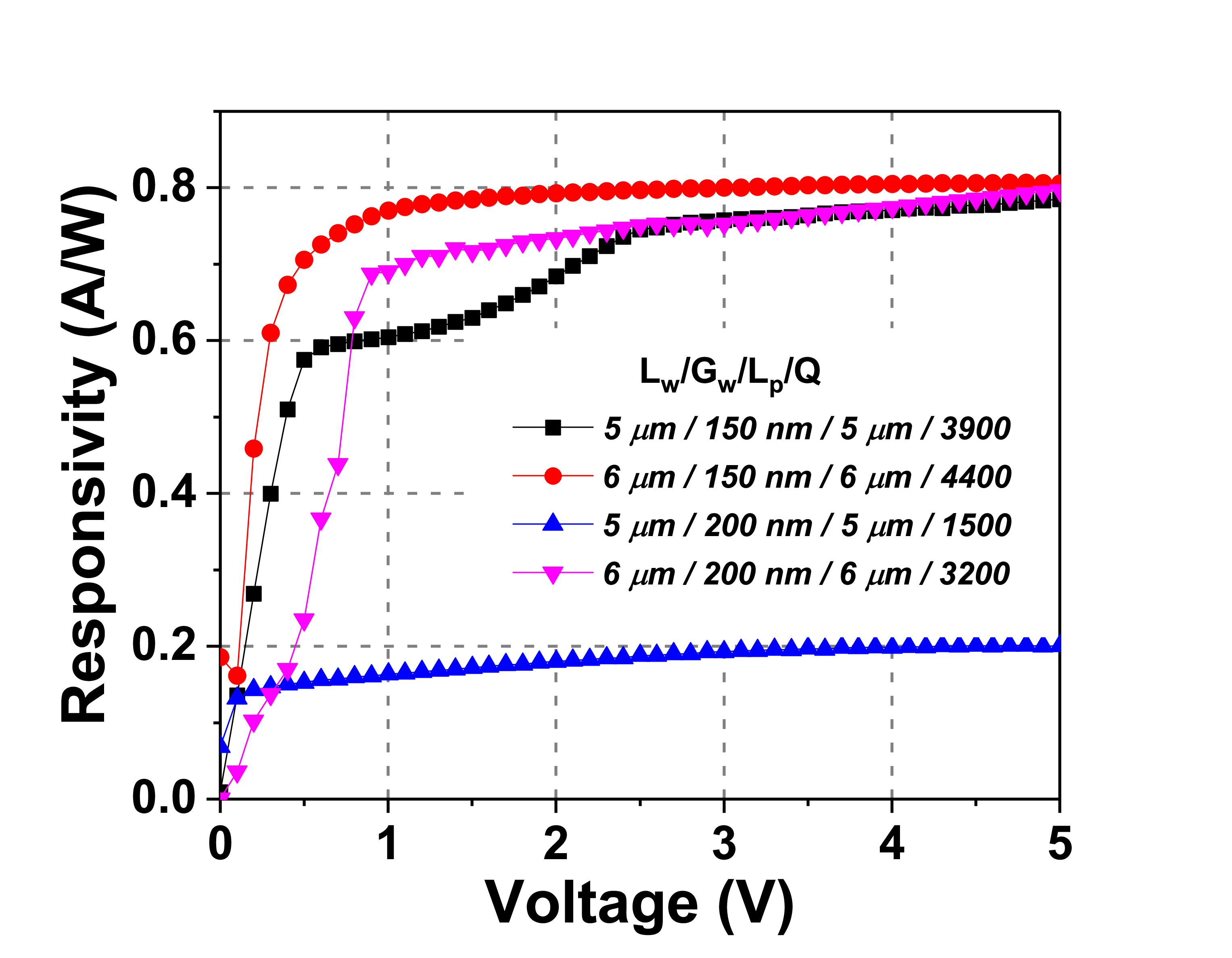}
\caption{Effect of detector length ($L_{p}$), ring coupling ($G_{w}L_{w}$), and Q-factor (Q) on the RCPD responsivity at various bias voltages.}
\label{fig:RvsringGL}
\end{figure}

The quality factor of the ring resonator has a direct influence on the detector responsivity. Figure \ref{fig:RvsringGL} shows the effect of light-coupling into the ring and the quality factor. For the devices with quality factor $>$3000, we observe higher responsivity ($\approx$0.75 A/W), however, for a device with lower coupling due to larger gap $G_{w}$ = 200 nm and shorter coupling length $L_{w}$ = 5$\mu$m (Fig. \ref{fig:directional_coupler}), we observe a lower responsivity of 0.2 A/W. Cavity enhancement is achieved with optimal coupling between the waveguide and ring that compensates loss in the cavity to achieve decent quality factor.  Table \ref{tab:msmcomp} summarizes the performance metrics of WGPD and RCPD with various power coupling. With increasing bias, we observe enhanced responsivity due to efficient charge collection; however, with a marginal increase in dark current. Furthermore, as mentioned earlier, the responsivity is low for cavities with lower quality factor.

\begin{table}
\centering
\caption{Performance comparison between waveguide coupled photodetector (WGPD) and ring coupled photodetector (RCPD) for different coupling gap $G_{w}$, coupling length $L_{w}$, and detector length $L_{p}$ as depicted in Fig. \ref{fig:model} and Fig. \ref{fig:wgpd}. }
\begin{tabular}{|c|c|c|c|c|c|c|} 
\hline
\multirow{3}{*}{Parameters}         & \multirow{3}{*}{\makecell{Bias\\ (V)}} & \multirow{3}{*}{\makecell{WGPD \\ $L_{p}$= 6 $\mu m$}  } & \multicolumn{4}{l|}{~ ~ ~ ~ ~ ~ ~ ~ ~ ~ ~ ~ ~ ~ ~ ~  ~ ~ ~ RCPD}                                         \\ 
\cline{4-7}
                                    &                       &                       & \multicolumn{2}{l|}{~ ~ ~ $G_{w}$ = 150 nm} & \multicolumn{2}{l|}{~ ~ ~ ~$G_{w}$ = 200 nm}  \\ 
\cline{4-7}
                                    &                       &                       & $L_{w}$= 5 $\mu m$ & $L_{w}$ = 6 $\mu m$                 & $L_{w}$ = 5 $\mu m$ & $L_{w}$ = 6 $\mu m$                    \\ 
\cline{4-7}
                                    &                       &                       & $L_{p}$= 5 $\mu m$ & $L_{p}$ = 6 $\mu m$                 & $L_{p}$ = 5 $\mu m$ & $L_{p}$ = 6 $\mu m$                    \\                                     
\hline
\multirow{3}{*}{\makecell{Dark Current\\ (nA)}}  & 1                     & 3.76                  & 3.00      & 3.76                       & 0.99      & 1.12                         \\ 
\cline{2-7}
                                    & 3                     & 6.39                  & 9.28      & 6.39                       & 1.41      & 5.23                         \\ 
\cline{2-7}
                                    & 5                     & 7.45                  & 13.1      & 7.45                       & 1.81      & 7.20                          \\ 
\hline
\multirow{3}{*}{\makecell{Photocurrent\\ (nA)}}  & 1                     & 49.9                  & 452       & 762                        & 486       & 472                          \\ 
\cline{2-7}
                                    & 3                     & 59.2                  & 566       & 792                        & 571       & 515                          \\ 
\cline{2-7}
                                    & 5                     & 66.6                  & 587       & 797                        & 596       & 545                          \\ 
\hline
\multirow{3}{*}{\makecell{Responsivity\\ (A/W)}} & 1                     & 0.005                 & 0.6       & 0.76                       & 0.16      & 0.69                         \\ 
\cline{2-7}
                                    & 3                     & 0.006                 & 0.76      & 0.79                       & 0.19      & 0.75                         \\ 
\cline{2-7}
                                    & 5                     & 0.007                & 0.78      & 0.80                     & 0.2       & 0.80                        \\
\hline
\end{tabular}
\label{tab:msmcomp}
\end{table}

\subsection{High-frequency response}

The frequency response of the Si-MSM photodetector was determined by the impulse response to a femtosecond optical pulse. Light from an 850 nm femtosecond pulsed laser source is coupled through the grating coupler. The full width at half maximum (FWHM) of 280 fs at 80 MHz repetition rate was used for the characterization. The light output from the pulsed laser source was focused vertically on top of the grating. The response of the photodetector was measured through a high-speed RF probe and sampling oscilloscope. A detailed schematic is presented in \cite{chatterjee2019high}.

\begin{figure}[ht]
\begin{subfigure}{0.45\textwidth}
\centering

\includegraphics[width=1\linewidth]{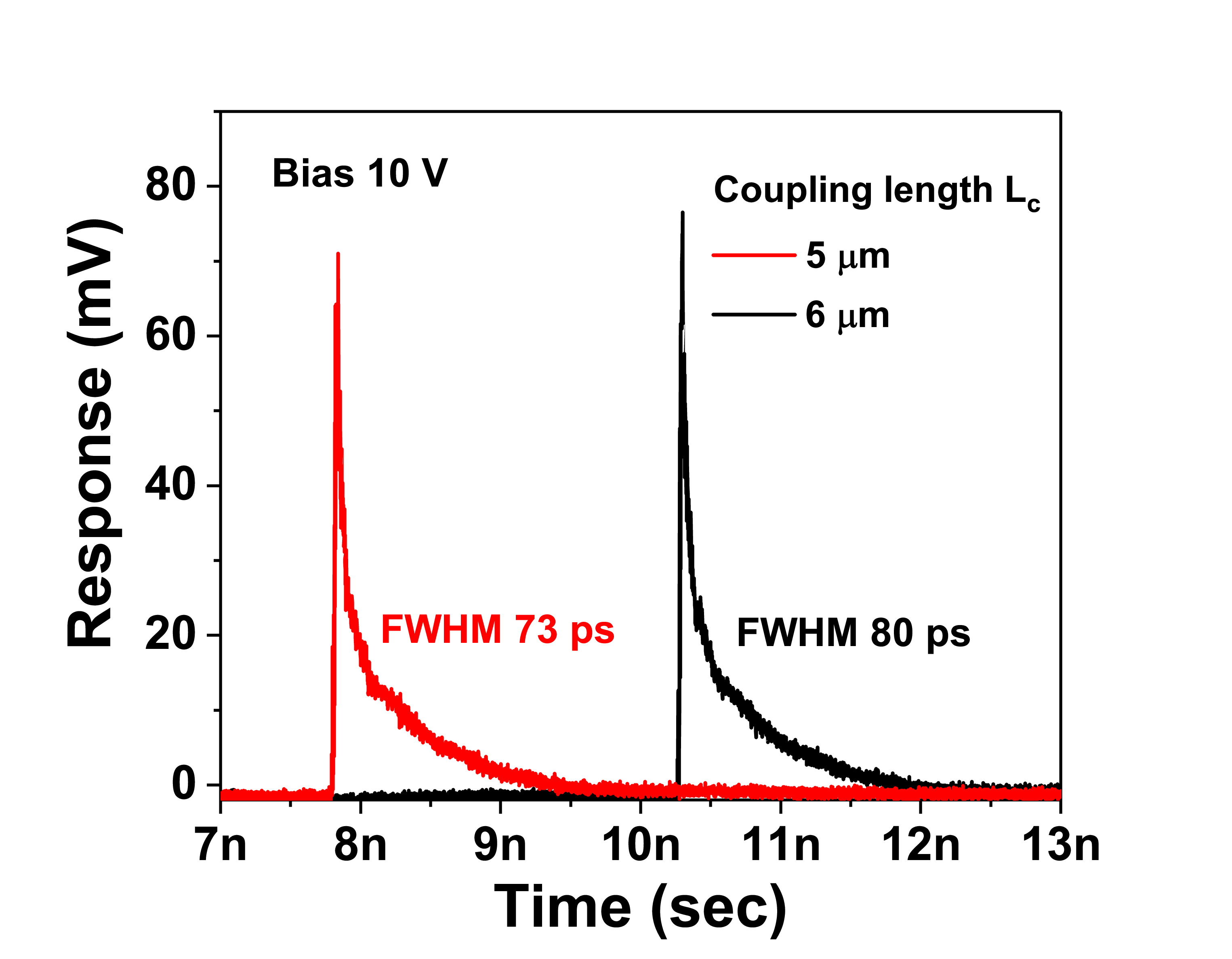} 
\vspace{-0.7cm}
\caption{Ring to bus coupling gap ($G_{w}$) 150 nm}

\end{subfigure}
\begin{subfigure}{0.45\textwidth}
\centering
\includegraphics[width=1\linewidth]{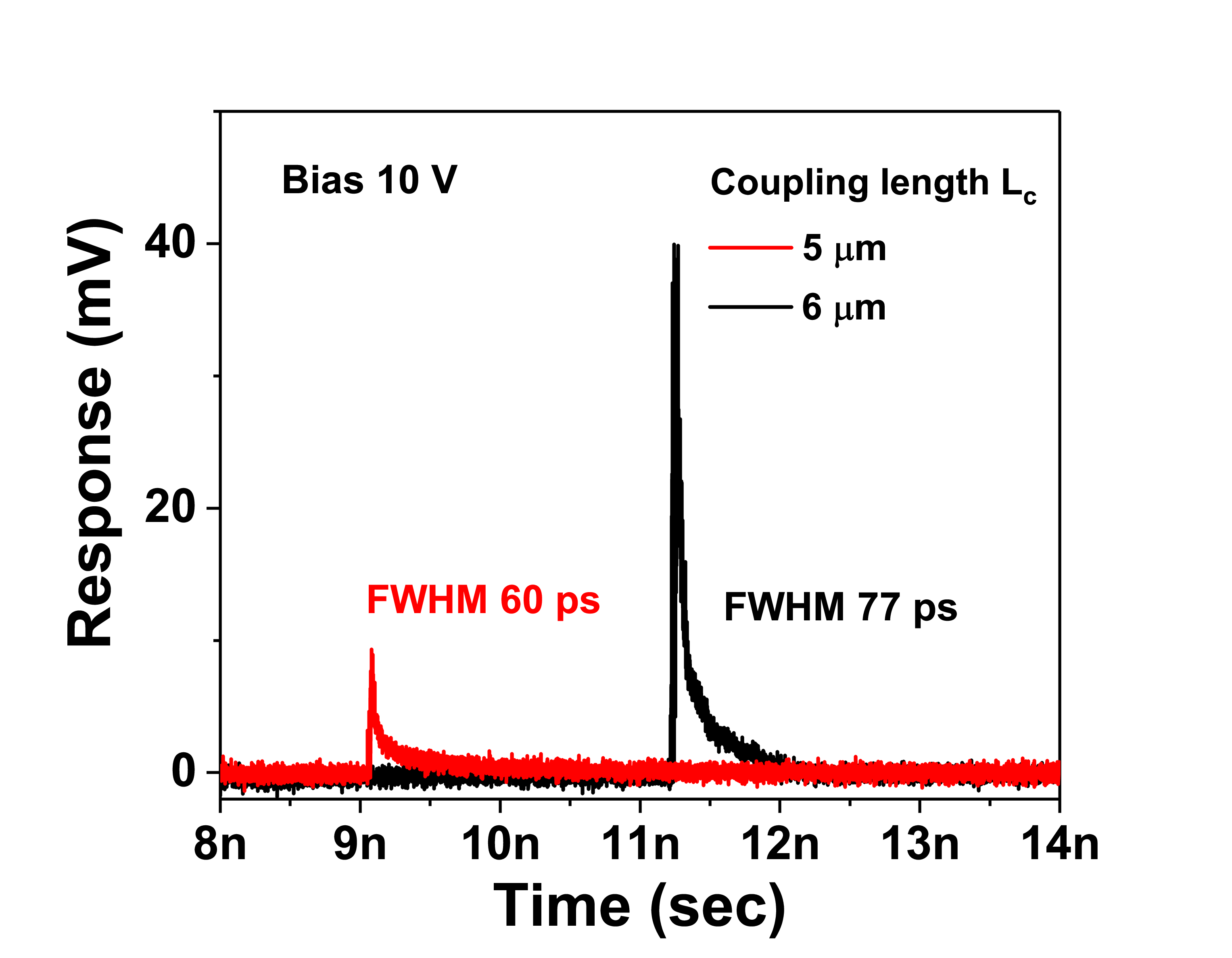}
\vspace{-0.7cm}
\caption{Ring to bus coupling gap ($G_{w}$) 200 nm} 
\end{subfigure}
\vspace{-0.3cm}
\caption{Femtosecond pulse response of ring coupled Si-MSM photodetector.}
\label{fig:pulserc}
\end{figure}

Fig \ref{fig:pulserc} (a) and (b) show the temporal response of the ring enhanced Si-MSM photodetector for different ring resonator parameters. The FWHM of the pulse response is used to determine the 3dB bandwidth of the detector as given in Eq. \ref{eq:f3db} \cite{Chen:09}.

\begin{equation}
    f_{3dB} = \frac{0.45}{FWHM}
    \label{eq:f3db}
\end{equation}

\begin{figure}[ht]
\centering
\includegraphics[width=0.5\linewidth]{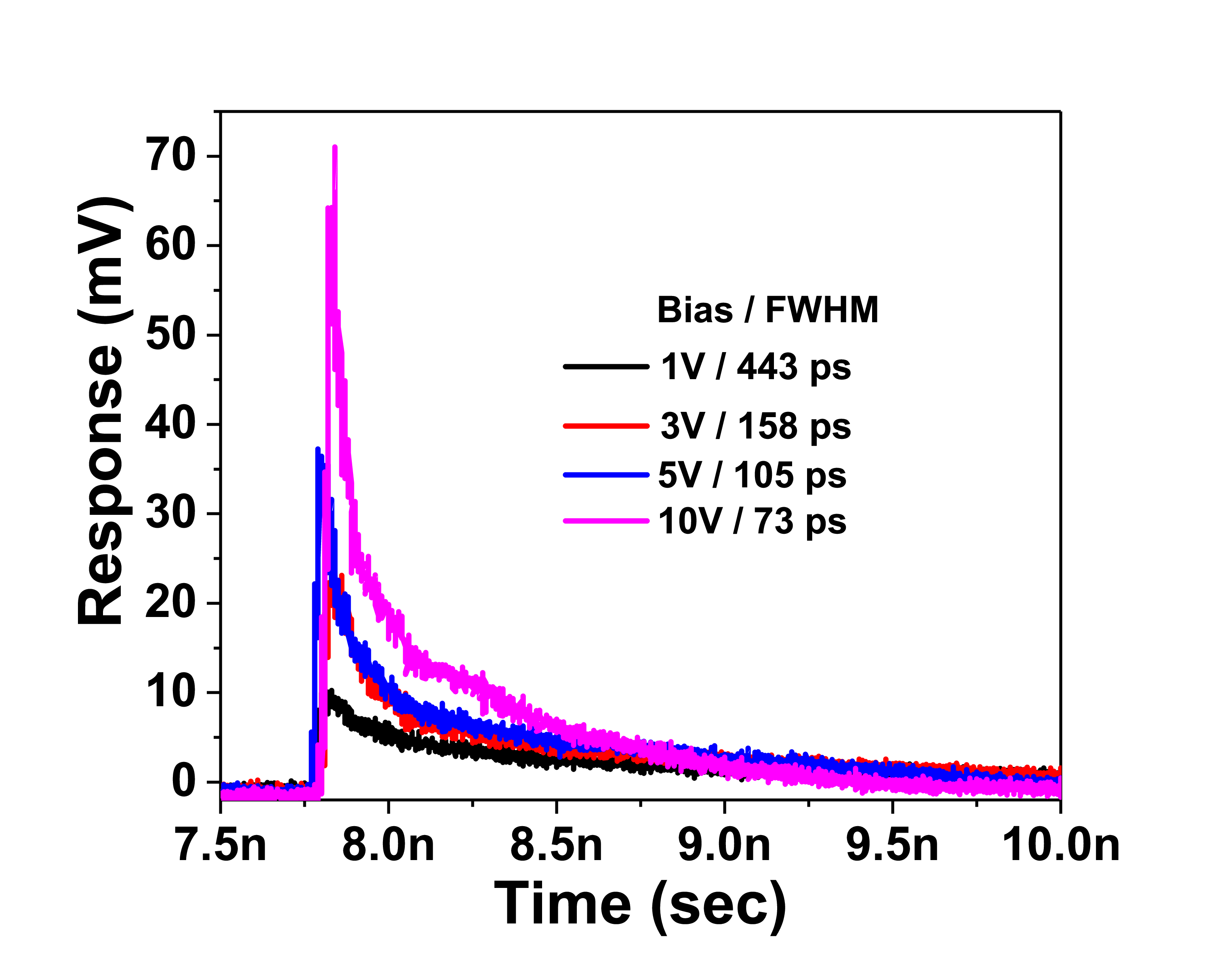} 
\vspace{-0.2cm}
\caption{ Effect of bias voltage on the pulse response of RCPD ($L_{p}$= 5 $\mu m$, $G_{w}$ = 150 nm, and $L_{w}$ =  5 $\mu m$)}
\label{fig:pulsercwg}
\end{figure}

\begin{figure}[!h]
\begin{subfigure}{0.45\textwidth}
\centering
\includegraphics[width=1\linewidth]{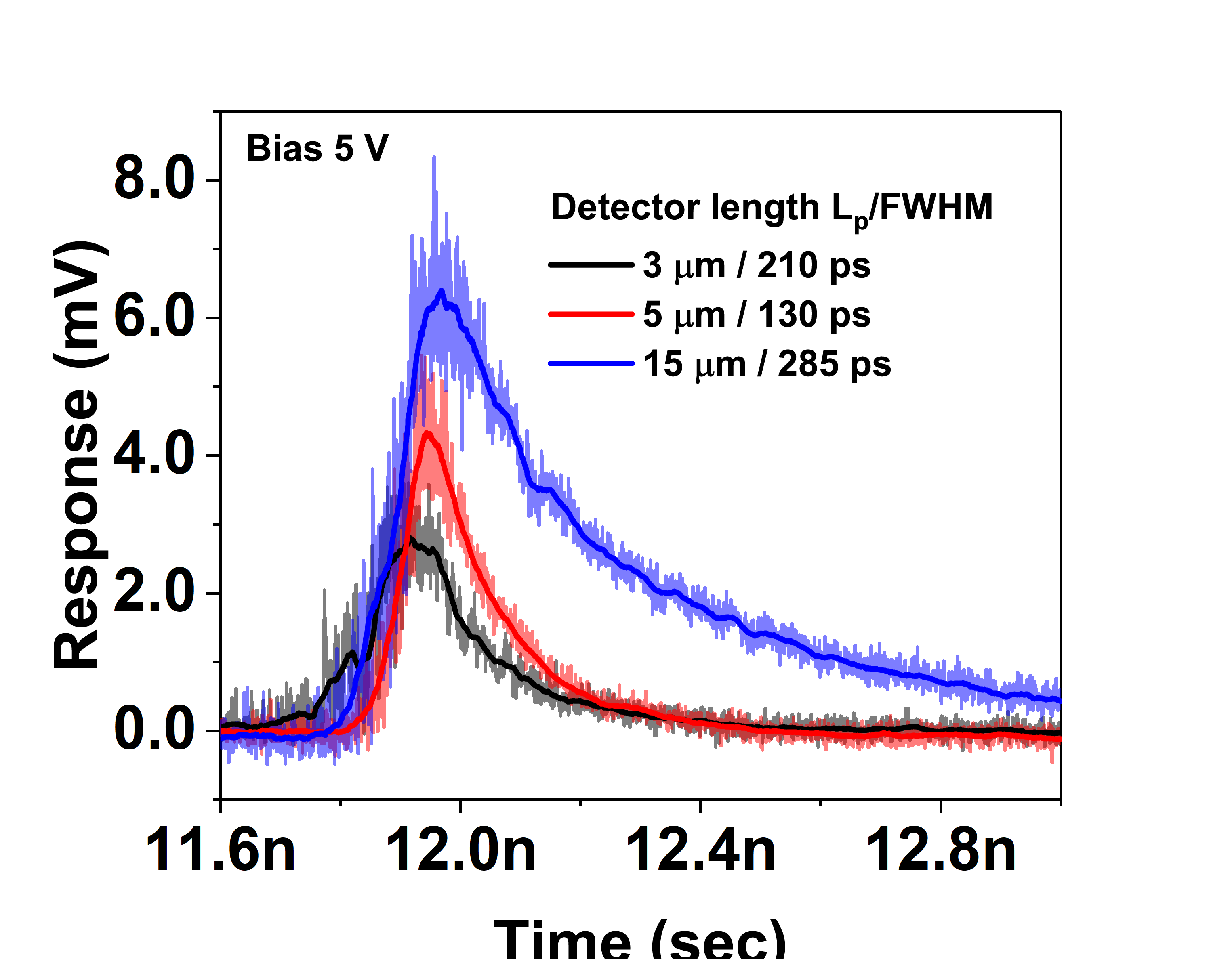} 
\vspace{-0.5cm}
\caption{Bias Voltage = 5 V}

\end{subfigure}
\begin{subfigure}{0.45\textwidth}
\centering
\includegraphics[width=1\linewidth]{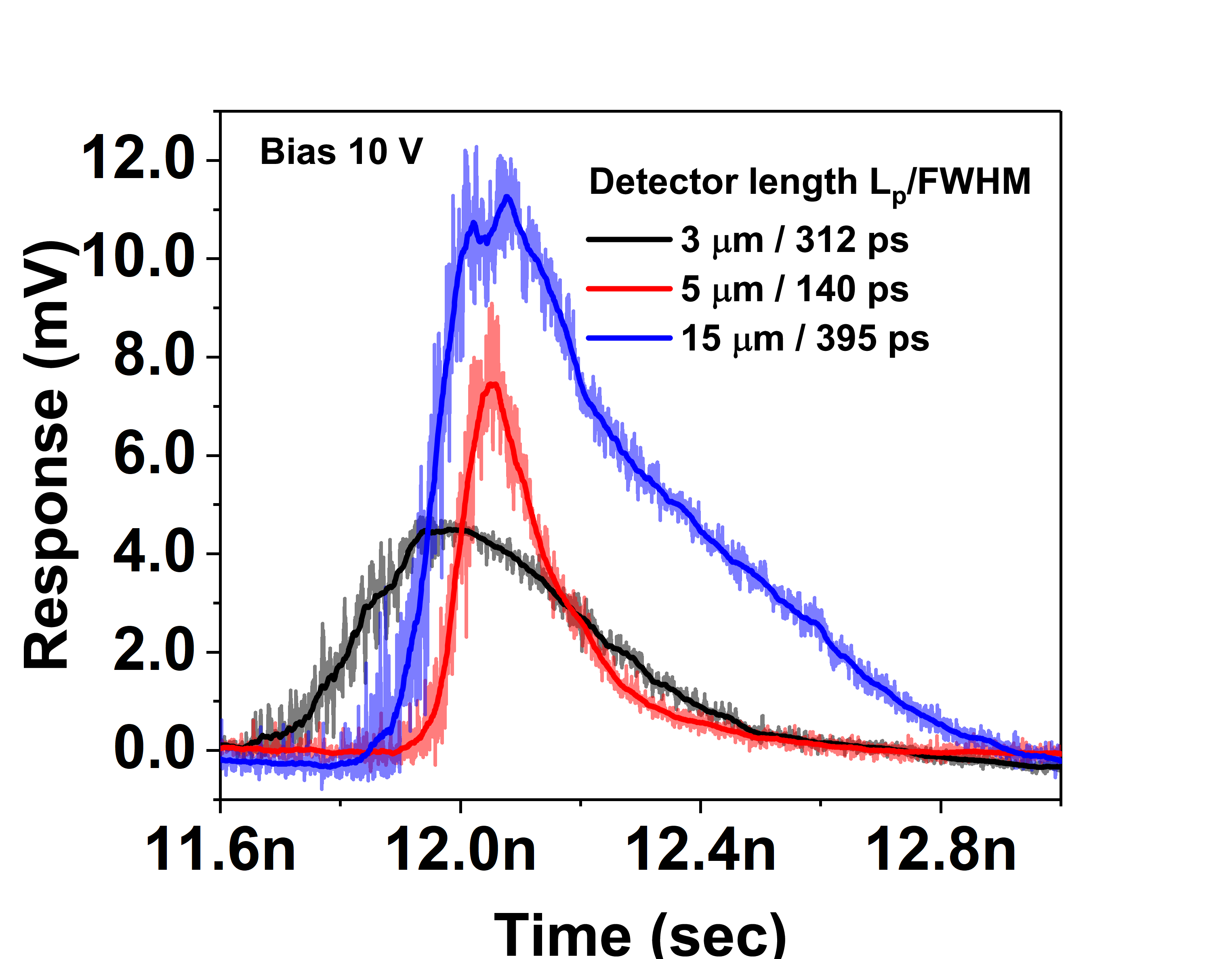}
\vspace{-0.5cm}
\caption{Bias Voltage = 10 V} 
\end{subfigure}
\vspace{-0.3cm}
\caption{Pulse response of waveguide coupled Si-MSM.}
\label{fig:pulsewg}
\end{figure}

\begin{table}[]
\centering
\caption{Comparison with various demonstration of photodetectors in 850 nm wavelength band}
\Large
\resizebox{\textwidth}{!}{%
\begin{tabular}{lclclclclclclclclcl}
\hline
\hline
Ref. & Bias (V) & \vtop{\hbox{\strut Dark current }\hbox{\strut (pA/$\mu m^{2}$)}} & R (mA/W)  & $f_{3dB} (GHz)$ & $\eta (mA/W * GHz)$ & \vtop{\hbox{\strut Platform }\hbox{\strut (Active layer thickness)}} & Electrode separation & Integrated & \vtop{\hbox{\strut Resonant cavity }\hbox{\strut enhancement factor}} \\ \hline
 \cite{liu1994140} & 1 & 0.2  & 5.7 & 140 & 798 & SOI, Si MSM (100 nm) & 100 nm & No & Nil\\ \hline
 \cite{samusenko2016sion} & 9 & $7.66X10^{-3}$  & 330 & Nil & Nil & Poly Si pin (150 nm) & NA & Yes & Nil\\ \hline
 \cite{PourFard2017} & 14 & $2.6X10^{3}$  & 300 & 16.4 & 4920 & SOI, Si pin (220 nm) & NA & No & Nil\\ \hline
\cite{gao2017photon} & 5 & $4X10^{-2}$  & 340 & 15 & 5100 & Micro structured Si pin (2 $\mu m$) & NA & No & Nil\\ \hline
 \cite{cheng2005silicon} & 4 & $4X10^{-3}$ & 230 & 0.96 & 221 & SOI, Si pin ($2.18\mu m$) & NA & No & 4.6X\\ \hline 
 \cite{schaub1999resonant} & 5 & $2.7X10^{-3}$ & 210 & 34 & 7140 & Si pin (500 nm) & NA & No & Nil\\ \hline 
 This work & 5 & 45 & 790 & 5 & 3950 & SiN-on-SOI, Si MSM (220 nm) & 4 $\mu m$ & Yes & 100X\\ \hline
 This work & 10 & 292 & 810 & 7.5 & 6075 & SiN-on-SOI, Si MSM (220 nm) & 4 $\mu m$ & Yes & 100X\\ \hline
 \hline
\end{tabular}%
}
\label{tab:my-table}
\end{table}

Figure  \ref{fig:pulsewg} depicts the pulse response of WGPD of various lengths at a bias of 5 V and 10 V. We measure best FWHM of 130 ps corresponds to a $f_{3dB}$ of 3.46$\pm$0.4 GHz for a detector length of 5 $\mu m$. It is evident from Figure  \ref{fig:pulsewg}a and Figure  \ref{fig:pulsewg}b that increasing the bias improves the responsivity; however, the pulse response time increases. Even for the same detector length of $5$ $\mu m$, at 5 V bias, we measure a pulse response time of 130 ps and 105 ps from WGPD and RCPD, respectively. We attribute the lower bandwidth of WGPD to charge screening effect due to photocurrent saturation \cite{lin1997high}. One could decrease the screening effect by reducing the power in the waveguide. However, lower input power would result in lower photocurrent due to poor responsivity of WGPD. Although increasing the length of the photodetector to 15 $\mu m$ improves the responsivity, the pulse response does not improve. The speed of longer detector is limited by transit time. Thus a trade-off between screening and transit time limit is enviable. A 5 $\mu m$ long WGPD shows such optimum where we observe response times are better than 3 $\mu m$ long device with higher responsivity.

Table \ref{tab:my-table} summarizes reported photodetector performance metrics in the literature. The ring coupled photodetector demonstrated has the highest responsivity-gain product ($\eta$). Although silicon pin photodetectors show higher $\eta$, MSM based detectors are still attractive due to potential higher bandwidth.


\section{Conclusion}
  
In summary, we have demonstrated SiN ring resonator enhanced silicon-MSM photodetector in the 850 nm wavelength band on SiN-SOI platform. Compared to waveguide photodetector, ring enhanced MSM with identical dimension gives two-orders enhancement in responsivity. Responsivity enhancement is attributed to the enhanced optical field inside the ring cavity and absorption of the optical field in the Si-MSM photodetector. As a result of field enhancement, we achieve compact photodetector with high responsivity. We demonstrate a responsivity of $0.81$ $A/W$ at $10$ $V$ with a $5$ $\mu m$ long detector. The smaller physical size of the photodetector also ensures low RC time constant and hence large bandwidth. With a 5 $\mu m$ long device, we achieved an electro-optic bandwidth of 7.5 GHz. We also presented a detailed discussed on the effect of device configuration on the photodetector response. In conclusion, on-chip wavelength-selective Si-MSM paves the way to realize a highly efficient, compact, and cost-effective solution for the application in optical interconnects and lab-on-chip bio-sensors in $<$1100 nm wavelength range.    

\section{Acknowledgement}
We acknowledge funding support from MHRD through NIEIN project, from MeitY and DST
through NNetRA project.



\end{document}